%% file: aaai2023.tex
\documentclass[letterpaper]{article} 
\usepackage{aaai23}  
\usepackage{times}  
\usepackage{helvet}  
\usepackage{courier}  
\usepackage[hyphens]{url}  
\usepackage{graphicx} 
\usepackage{natbib}  
\usepackage{caption} 
\usepackage{amsmath}
\usepackage{listings}
\usepackage{multirow}
\usepackage{soul}
\usepackage{comment}
\usepackage{amsmath}
\usepackage{amssymb}
\usepackage{amsthm}
\usepackage{xcolor}
\usepackage{filecontents}
\usepackage{subfigure, hyperref}
\usepackage{multirow,diagbox}
\usepackage[ruled,noline,noend,linesnumbered]{algorithm2e}
\usepackage{newfloat}
\usepackage{listings}

\newcommand{\subhead}[1]{\vspace {0.05in}\noindent{\textbf{#1.}}}
\DeclareCaptionStyle{ruled}{labelfont=normalfont,labelsep=colon,strut=off} 
\usepackage{pifont}
\newcommand{\cmark}{\ding{51}}%
\newcommand{\xmark}{\ding{55}}%

\def \draft {1}

\if \draft 1

\else

\fi

\title{Label Inference Attack against Split Learning under Regression Setting}
\author{
    Shangyu Xie,\textsuperscript{\rm 1} 
    Xin Yang,\textsuperscript{\rm 2} 
    Yuanshun Yao,\textsuperscript{\rm 2} 
    Tianyi Liu,\textsuperscript{\rm 2}
    Taiqing Wang,\textsuperscript{\rm 2} 
    Jiankai Sun \textsuperscript{\rm 2}\\
}
\affiliations{
    \textsuperscript{\rm 1} {Illinois Institute of Technology}, 
    \textsuperscript{\rm 2} {ByteDance Inc.}\\

    sxie14@hawk.iit.edu, \{yangxin.yx, kevin.yao, tianyi.liu, taiqing.wang, jiankai.sun\}@bytedance.com
}

\usepackage{bibentry}

\begin{document}

\maketitle

\begin{abstract}

As a crucial building block in vertical Federated Learning (vFL), Split Learning (SL) has demonstrated its practice in the two-party model training collaboration, where one party holds the features of data samples and another party holds the corresponding labels. Such method is claimed to be private considering the shared information is only the embedding vectors and gradients instead of private raw data and labels. However, some recent works have shown that the private labels could be leaked by the gradients. These existing attack only works under the classification setting where the private labels are discrete. 
In this work, we step further to study the leakage in the scenario of the regression model, where the private labels are continuous numbers (instead of discrete labels in classification). This makes previous attacks harder to infer the continuous labels due to the unbounded output range. To address the limitation, we propose a novel learning-based attack that integrates gradient information and extra learning regularization objectives in aspects of model training properties, which can infer the labels under regression settings effectively. The comprehensive experiments on various datasets and models have demonstrated the effectiveness of our proposed attack. We hope our work can pave the way for future analyses that make the vFL framework more secure. We
release code for future work and reference\footnote{\url{https://github.com/xiehahha/aaai_ppai23_split_learning_leakage}}.

\end{abstract}

\section{Introduction}

\input{intro}


\section{Background \& Related Work}

\input{background}

\section{Proposed Attack}

\input{attack}

\section{Experimental Evaluation}
\input{exp.tex}

\subsection{Defense Evaluation}
\input{dis.tex}

\section{Conclusion}

In this paper, we have investigated the privacy risks of Split Learning with regression problems. We propose a novel learning-based attack with multiple regularizations to infer the labels with high accuracy. We have conducted comprehensive experiments to demonstrate the effectiveness of the proposed attack and also present the insights on the attack results correspondingly. We believe that the proposed attack can motivate more defense work in this field.

\end{document}

%% file: intro.tex
\begin{figure}[!th]
	\centering{
		\includegraphics[angle=0, width=0.85\linewidth]{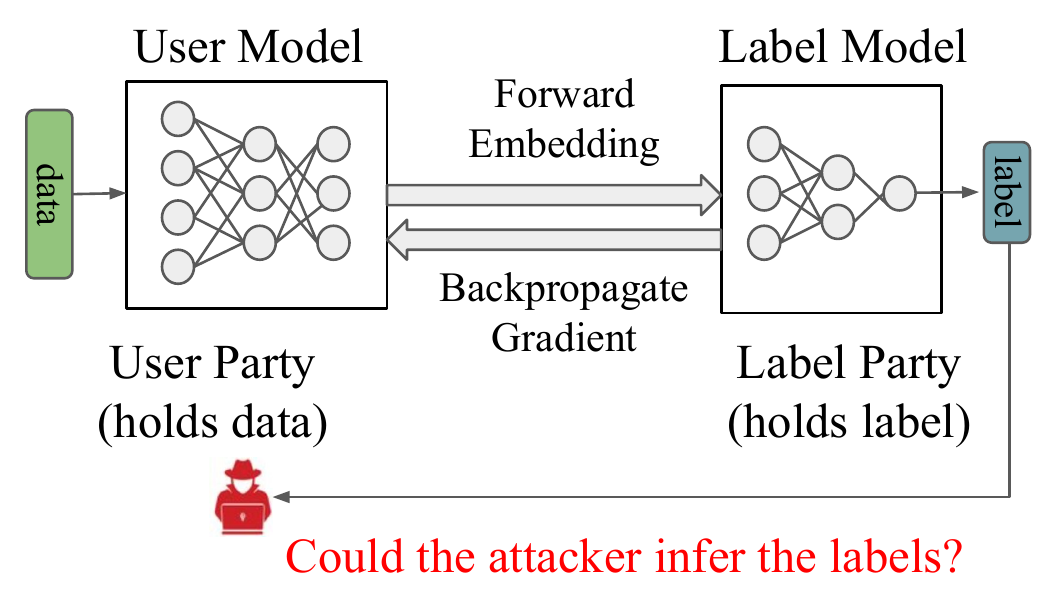}}
	\caption[Optional caption for list of figures]
	{Privacy Attack in Split Learning. Different from previous attacks on inferring \emph{discrete} labels for classification, we study the \emph{continuous} value with regression problem.}
	\label{fig:att_f}
\end{figure}

With the rising privacy concerns in Machine Learning, Split Learning (SL) \cite{gupta2018distributed, vepakomma2018split,li2022label,thapa2022splitfed} techniques have emerged as one of main privacy-enhancing techniques (PETs), which have been shown their practice in online advertisement conversion prediction tasks \cite{sun2021defending}. Specifically, as a fundamental building block for vertical Federated Learning (vFL) \cite{feng2020multi, chen2020vafl,liu2020asymmetrical, yang2019federated}, Split Learning allows two parties (``user party'' holding data inputs and ``label party'' holding labels, respectively) to jointly train a machine learning model (``user model'' and ``label model'') without disclosing either original data inputs or labels. The only shared information is the intermediate layer embedding (for forward computation) or gradients (for backforward update), which seems to protect both data and labels at first glance. Figure \ref{fig:att_f} overviews the learning process among two parties.

Unfortunately, no rigorous proofs have shown that gradients or intermediate embedding of Split Learning can prevent data leakage. Additionally, some recently proposed privacy attacks \cite{erdogan2021unsplit,li2022label,fu2022label,sun2021defending} have demonstrated that the sensitive labels can be inferred by a curious user party with very high accuracy in aspects of classification problem. For example, Unsplit \cite{erdogan2021unsplit} proposes a gradient-matching based attack which minimizes the MSE distance between the original gradient and surrogate's to infer labels. Another representative work \cite{li2022label} utilizes both directional and sign (positive/negative) difference between the gradients of binary labels to effectively infer the original labels. Fu et al. \cite{fu2022label} propose a semi-supervised learning-based method to train a surrogate model with some auxiliary labeled data samples (and thus infer the labels).

However, current privacy attacks against Split Learning have the following limitations. On the one hand, almost all the attacks only study the classifications model \cite{erdogan2021unsplit,li2022label,fu2022label} on various discrete-labeled datasets. For example, practical advertisement conversion prediction is a binary classification \cite{li2022label} and image classifications (e.g., CIFAR10/100) are also popular with privacy attack evaluation for Split Learning \cite{fu2022label, erdogan2021unsplit}. Thus it would be essential to study the leakage of regression models considering regression is also popular among various applications, such as advertisement revenue prediction \cite{wurfel2021online,cikm22} and medical admission days prediction \cite{medical2,medical1}. One similar example to advertisement conversion is the prediction of advertisement revenue, which could potentially work with Split Learning framework \cite{wurfel2021online}. 

On the other hand, previous attacks on inferring discrete classification labels can  neither be directly adopted for regression problems nor applicable to practical attack scenarios (e.g., black-box or too much data knowledge). For example, DLG \cite{zhu2019deep} is the first work to demonstrate the leakages from gradients in white-box setting. Li et al. \cite{li2022label} mainly utilize the differences between the binary labels to infer the labels with advertisement conversion problem. Another representative attack \cite{fu2022label} relies on the a set of auxiliary data, of which the size influence the attack performance greatly. \cite{erdogan2021unsplit} works only with  label models of very few layers, e.g., 1-layer, which cannot be applied to more-layer neural networks. Table \ref{tab:comp} summarizes the main properties of existing attacks and ours. Note that \cite{fu2022label}'s performance would rely heavily on the size of auxiliary dataset. For instance, the attack performance would drop greatly from $0.8544$ to $0.6554$ if the size of auxiliary dataset reduces from $320$ (6.4\%) to $10$ (0.2\%).

\begin{table}
  \centering
  \scriptsize
   \caption{Comparison of our work and existing label inference attacks against split learning. Our proposed attack focuses on the regression problem under the black-box with only very few labeled data ($<\sim1\%$ out of whole dataset).}
  \label{tab:com}
  \begin{tabular}{c|ccc}
    \hline
    Method &Black-box & Regression & Auxil. Knowledge\\\hline
    DLG & \xmark & \xmark & Full model\\
    \cite{li2022label} & \cmark & \xmark &Imbalanced data\\
    \cite{fu2022label} & \cmark& \xmark & Large ($>\sim5\%$) labeled data\\\hline
    Ours & \cmark & \cmark & Few ($<\sim1\%$) labeled data\\\hline
    
  \end{tabular}

\label{tab:comp}
\end{table}

To this end, we propose a novel privacy attack to infer the labels via a learning-based approach with carefully designed learning objectives, which mainly utilize the shared back-propagate gradient information and other learning regularizations to improve the attack performance. Specifically, under the black-box setting, the attacker will try to utilize a surrogate model and dummy label to recover the original label value. The attacker would try to match the gradients returned by the surrogate model with the original gradients (by minimizing the distances of gradients). Besides, we add several learning regularizations which can help the attack optimization to converge to the original value. First is the normal model training performance, i.e., the surrogate model's prediction should be close to the dummy label since the training process will enforce the model to fit with the dataset. Another is semi-supervised learning with some auxiliary labeled data. We can further restrict the optimization of the surrogate model and label by directly utilizing such labeled data information. We have experimentally validated the effectiveness of the newly proposed attack on several datasets. To summarize, we make the following contributions.

\begin{itemize}
    \item To the best of our knowledge, we are the first to evaluate \emph{black-box} leakage against Split Learning under the regression problem.
    \item We propose a learning-based privacy attack that can infer the labels with carefully designed learning objectives based on gradient and regularizations of both model and data properties. 
    \item We have conducted extensive experiments to evaluate the proposed attack on several datasets and the results have demonstrated the effectiveness of the attack. We have also evaluated some commonly-used defense schemes against the proposed attack, which would motivate more future work.  
\end{itemize}

%% file: background.tex
\begin{figure*}[!th]
\centering{
\includegraphics[angle=0, width=0.88\linewidth]{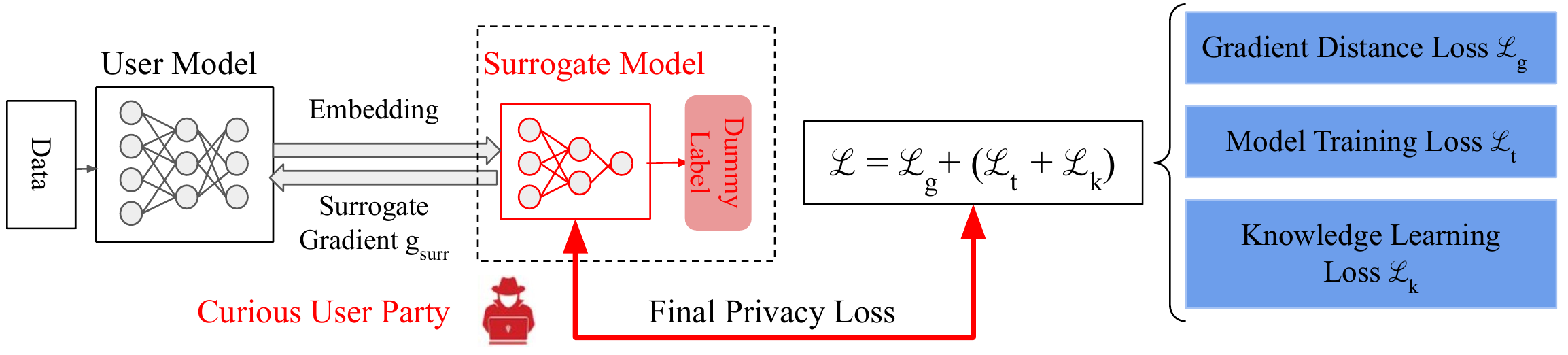}}
\caption{Overview of attack framework. The attacker (curious user party) which wants to infer the corresponding labels from the label party will first set up the surrogate model and dummy label. The attacker will try to update both the surrogate model and labels with the privacy loss, which includes gradient distance loss $\mathcal{L}_{g}$ and other two regularization losses $\mathcal{L}_{t}$ and $\mathcal{L}_{k}$.}
\label{fig:att_framew}
\end{figure*}

\subsection{Two-party Split Learning in FL}

Split Learning \cite{gupta2018distributed, vepakomma2018split} has been proposed to enable two parties (user party and label party) to collaboratively train a composite model with the vertically partitioned data (usually in Federated learning \cite{yang2019federated}). Such composite model is split into user model and label model held by user and label party respectively. More formally, for user party, denote the user's input data set $X=\{x_i, i\in [1, |D|]\}$, and the user model $\mathcal{M}_{user}$; for the label party, denote the label party's label $Y=\{y_i, i\in [1, |D|]\} \in \mathbb{R}^1$ (corresponding to $x_i$) and label model $\mathcal{M}_{label}$. $D=\{(x_i, y_i)\}$ is the whole training set to use for training the composite model $\mathcal{M}_{user} \circ \mathcal{M}_{label}$. 

Following the conventional training process, Split Learning also consists of two phases: 1) forward pass and 2) backpropagation. 

\subhead{Forward pass} The user party will compute the intermediate layer (namely cut layer) embedding $\mathcal{E}_i=M_{user}(x_i)$. This embedding will be used as input to the label model to compute the final output (prediction score) of the whole composite model:

\begin{equation}
    y_i^{predict}=\mathcal{M}_{label}(\mathcal{M}_{user}(x_i))=\mathcal{M}_{label}(\mathcal{E}_i)
\end{equation}
Then we can derive the loss for such prediction $y_i^{predict}$ as $L(y_i^{predict}, y_i)$.

\subhead{Backpropagation} The label party will update its label model's parameters by computing the gradient of loss $L$ with respect to the label model itself $\mathcal{M}_{label}$. To enable the user party to compute the updates for user model $\mathcal{M}_{user}$, the label party will need to compute the \emph{shared} gradient with respect to the cut layer embedding $\mathcal{E}_i$ by chain rule, $\nabla_{\mathcal{E}_i}L$ (denoted as $g$) as following: 

\begin{equation}
    \nabla_{\mathcal{E}_i}L=\frac{\partial L(y_i^{predict}, y_i)}{\partial y_i^{predict}}\cdot \frac{\partial y_i^{predict}}{\mathcal{E}_i}
    \label{eq:embedding}
\end{equation}

Then the user party can compute the gradient update for the user model w.r.t. $\mathcal{M}_{user}$'s parameters:

\begin{equation}
    \nabla_{\mathcal{M}_{user}}L=\nabla_{\mathcal{E}_i}L \cdot \frac{\partial \mathcal{E}_i}{\mathcal{M}_{user}}
\end{equation}

As for the inference phase, the user party will first compute the cut layer embedding $\mathcal{E}_i=\mathcal{M}_{user}(x_i)$ and send it to the label party, which will compute the final prediction. 

\subsection{Related Work}

\subhead{Privacy Attacks against Split Learning} The label leakage under Split Learning has been first studied in the context of advertisement conversion prediction \cite{li2022label}. The proposed attack utilizes the difference of shared gradients (including both magnitude and direction) between the positive and negative data samples, which achieve a high AUC score for inferring labels. However, such attack relies much on the imbalance of the dataset and only focuses on binary classification \cite{li2022label}. Another work model completion attack \cite{fu2022label} adopts semi-supervised learning to train a surrogate model with the auxiliary dataset, which heavily relies on the size of the auxiliary dataset to achieve good attack performance. Such direct semi-supervised learning method gets much worse performance (as depicted in the experiments later). Our work is similar to the classic gradient matching \cite{zhu2019deep}, which utilizes a learning-based method to infer the data and labels by minimizing the distance of gradients under the white-box setting (known model). But our work has a stronger attack setting to focus on the \emph{black-box} setting, where the attacker does not know the target model.

Above all, all the previous works mainly utilize the difference of the gradients for the binary or multi-classification problem, which are not fit for the regression problem (the output score are continuous in a specific range). To our best knowledge, we are the first to study the leakage against the split learning under the regression problem. 

\subhead{Privacy-enhancing Technologies (PETs) in FL} There are two main categories to protect the data in ML: 1) carefully designed cyrptographic protocols, e.g., Secure Multiparty \cite{wagh2020falcon, mohassel2018aby3} or Two-party computation \cite{mohassel2017secureml,patra2021aby2,xie2021generalized}; 2) perturbation-based methods to obfuscate the shared information among the parties or directly perturb information to be protected \cite{abadi2016deep,erlingsson2019amplification,ghazi2021deep, sun2022dpauc,yang2022differentially, xie-hong-2022-differentially}. For example, labels are protected by randomized responses to satisfy label differential privacy \cite{ghazi2021deep}. \cite{sun2022dpauc} works on adding noise to the gradients to protect the data.

%% file: attack.tex
\subsection{Threat Model}

We define the threat model for our attack in aspects of security research, including the attack setting \& goal, the attacker's capability, and knowledge. 

\subhead{Attack Setting \& Goal} We follow the previous works \cite{li2022label, fu2022label} to study a semi-honest model, where the attacker (user party) will follow the normal training protocol but will try to infer the private labels of the label party corresponding to her own inputs. Note that we try to infer a score-based label in a regression problem instead of the classification problem. Our attack works under the black-box setting, i.e., the attacker does not know the label model architecture or parameters on the label party's side.

\subhead{Attacker's Capability \& Knowledge} For the black-box setting, the attacker would have the ability to design its own surrogate model (e.g., increasing the depth or width of the model) to attack. In addition, we also assume that the attack may hold a small set of data points with known labels, which will be taken as a training dataset to train the model \cite{fu2022label}. The attacker can achieve this by training with its local labeled dataset or artificial dataset during the normal Split Learning process. This is a reasonable setting since the label party as a service provider cannot determine if the data from the label party are artificial or not.

\subsection{Attack Method}

Our proposed attack is to get the private labels by a learning-based approach. Specifically, we will first construct a surrogate model for the label model and also the dummy labels we want to infer considering we are considering black-box attack. Then we will carefully define the loss function based on the known information (e.g., gradients) and other machine learning related objective (e.g., model performance and semi-supervised learning) as regularization. Then we will try to update the labels and model to get an optimal solution, which can be viewed as the original label. Our attack framework is demonstrated in Figure \ref{fig:att_framew}.

More formally, we give the attack notation for sake of demonstration. Denote the surrogate model and dummy labels corresponding to data $x_i$ set by the user as $\mathcal{M}_{surr}$ and $y_i^{dummy}$, respectively. Then we can compute both shared gradient (denoted as $g_{surr}$) and predictions of surrogate model following the normal Split Learning process.

\subhead{Gradient Distance Loss} The shared gradient for backpropagation (\autoref{eq:embedding}) is the main information we utilize to learn the private labels, i.e., the gradient $g_{surr}$ returned by the surrogate model $\mathcal{M}_{surr}$ must be close to the original $g$. Then we define a gradient distance loss which represents the distance between the original gradient $g$ and the surrogate gradient $g_{surr}$ computed with the surrogate model and dummy label:

\begin{equation}
    L_{g}=\mathcal{D}(g, g_{surr})
\label{eq:grad}
\end{equation}

where $\mathcal{D}(\cdot)$ is a distance function, e.g., L2 norm function. Our experimental evaluation also shows that such gradient distance loss can work with a batch of data, which can be easily scaled to large datasets.

\subhead{Training Accuracy Loss} Considering that our learning objective involves high-dimensional spaces, it could be very likely to get multiple solutions even the gradient distance loss is 0. Then to restrict our learning space and get the correct direction to the true label, we add an accuracy loss as regularization to bind the surrogate model to behave like a normal label model (which then tries to achieve high performance for the tasks). This is based on a mild assumption that we will target the label model during the latter stage of training, where the label model can fit well with the training dataset. To this end, we define the training performance loss as the predictions by the surrogate model $\mathcal{M}_{surr}$ should be close to the dummy labels $y_i^{dummy}$ when it is converged.

\begin{equation}
    L_{t}=||\mathcal{M}_{surr}(\mathcal{M}_{user}(x_i))-y_i^{dummy}||^2
\label{eq:acc_loss}
\end{equation}

We use the square of absolute value as the accuracy loss. 

\subhead{Knowledge Learning Loss} As mentioned in the threat model, we also assume that the attacker can hold a small set of dataset (the data with known labels) from the training set, denote as $\mathcal{D}_{known}=\{(x_j, y_j)\}$. We will aggregate the loss with both gradient distance loss and accuracy loss similarly but with the ground truth label $y_j$ instead of dummy label $y_i^{dummy}$. Combining with \autoref{eq:grad} and \autoref{eq:acc_loss}, we define the knowledge loss as:

\begin{equation}
    L_{k}=\sum_{(x_j, y_j)\in D_{known}}(L_{g}+L_{t})
\end{equation}

Besides, we also add the weight parameter $\lambda_1$ and $\lambda_2$ to balance the learning loss. Thus the overall learning loss function of our attack is as following:

\begin{equation}
    L_{loss}=L_{g}+ \lambda_1 \cdot L_{t}+\lambda_2 \cdot L_{k}
    \label{eq:total}
\end{equation}

We can utilize gradient-based algorithms, e.g., Adam \cite{kingma2014adam} to iteratively update both $\mathcal{M}_{surr}$ and $y_i^{dummy}$ (the dummy label could be solved with training batch).  Algorithm \ref{alg:dlg} demonstrates the optimization process. 
 
\begin{algorithm}

\KwIn{A batch of data $X=\{x_k, k\in[1, |B|]\}$\\
~~~~~~~~~~~~User party's model $\mathcal{M}_{user}$ \\
~~~~~~~~~~~~Shared Gradient from label party $g$\\
~~~~~~~~~~~~Number of Iterations $K$, Learning Rate $\eta$
}

\KwOut{Inferred label $Y^{\star}=\{y_k, k \in [1, |B|\}$ (w.r.t to every $x_k \in X$)}

\tcp{Construct initial dummy labels}

$\hat{Y}^{(0)}\gets_s [\mathcal{N}(0, 1)]^{|B|}$

Initialize the surrogate label model $\mathcal{M}_{surr}^{(0)}$

\For {$i \to 0$ to $K$}{

    \tcp{Attack loss per \autoref{eq:total} }
    $L_{loss}\gets \mathcal{F}(\hat{Y}^{(i)}; \mathcal{M}_{surr}^{(i)})$ 
    
    \tcp{Update $\hat{Y}^{(i)}$ and $\mathcal{M}_{surr}^{(i)}$}
    $\hat{Y}^{(i+1)}\gets \hat{Y}^{(i)}-\eta \nabla_{\hat{Y}^{(i)}}L_{loss}$\\
    $\mathcal{M}_{surr}^{(i+1)}\gets \mathcal{M}_{surr}^{(i)}-\eta \nabla_{\mathcal{M}_{surr}^{(i)}}L_{loss}$

}

\Return $\hat{Y}^{(K+1)}$ as ground-truth score

\caption{Proposed Learning-based Attack}\label{alg:dlg}
\end{algorithm}

%% file: exp.tex
\subsection{Experimental Setup}
\subhead{Datasets \& Models} We select the three different datasets on the regression problem: 1) Boston housing price \footnote{http://lib.stat.cmu.edu/datasets/boston}; 2) Power plant energy dataset \cite{energydataset}; 3) California housing price. Both Boston and California housing price dataset are used to predict housing price in Boston or California based on a set of properties while Power plant energy dataset is to predict the full-load electricity generated by power plants. \autoref{tab:data} summarizes the datasets and corresponding user/label models (``FC-3" denote as 3-layer Fully Connected neural network). 

\begin{table}[!h]
\centering
\caption{Dataset and Models}
\begin{tabular}{ccc}
\hline
{Dataset} &User Model & Label Model \\\hline

Boston Housing & FC-3 & FC-3\\
California Housing & FC-2 & FC-2\\
Energy  & FC-4 & FC-3 \\\hline

\hline

\end{tabular}
\label{tab:data}
\end{table}

\begin{table*}[!th]
\centering
\caption{Attack performance of proposed attack and baseline. We include the range of private labels for each dataset and the number of auxiliary known data to attack. We report both average L1 loss (ALV) and error rate (AER) for both methods.}
\begin{tabular}{cccccc}
\hline
{Dataset} & Dataset Size & Label Stats. (Min/Max) & Number of Known Data & Ours & Baseline\\\hline

Boston Housing & 400 & 5.0/50.0 & 4 & 2.31/3.47\%& 7.54/78.31\%\\

Energy Prediction & 6000 & 420.26/495.76 & 20 & 25.73/5.35\% &270.36/67.52\%\\

California Housing & 16000 & 0.15/5.0 & 40 & 0.11/4.16\%& 2.64/89.24\%\\\hline

\hline

\end{tabular}
\label{tab:main_attack}
\end{table*}

For every dataset's training, we use 80:20 split and the L1 loss as loss function for the regression problem.  W.l.o.g, we utilize the well-trained model (with a larger number of training epochs 15) as target model since the model is usually overfitted with the data under the distributed learning. All the trained models achieve good performance on the test dataset. For Boston housing dataset, the L1-loss is 2.24, Energy dataset is 21.52. and California is 0.08. For the surrogate model, we use FC-2 for the main attack experiments. We utilize Pytorch to implement DNNs and attack optimization. For the weight parameters, we tune $\lambda_1=1$ and $\lambda_2=0.005$. We set the size of learning batch to be $5$. We select the Adam as optimizer with learning rate $0.005$ and the number of iterations are 2000. Our attack evaluation will follow the same setting as depicted above unless stated. We repeat every single experiment 5 times and report the average value as final results.

\subhead{Attack Metrics} For regression problem, we define average L1-norm loss value (ALV) and average error rate (AER) to evaluate the attack performance:

\begin{equation}
    \textit{ALV}=\frac{\sum_{y_i \in Y} abs(y_i^\star-y_i)}{|Y|}
\end{equation}

\begin{equation}
    \textit{AER}=\frac{\sum_{y_i \in Y} abs(y_i^\star-y_i)/y_i}{|Y|}
\end{equation}

\subhead{Baselines} We first select the semi-supervised learning to infer labels as a benchmark, i.e., the attacker would directly utilize the known auxiliary data points to directly train the surrogate model, and use the surrogate model to infer target labels. A difference here is to only tune the surrogate model while fixing the user model (on attacker's side) instead of retraining the whole composition model since we only have a few data points (similar to few-shot learning).

\subhead{Evaluations} We would like to answer the following questions through the evaluation.

\begin{enumerate}
    \item Does the proposed attack work well with general dataset under regression? 
    \item Which factors in the attack  will impact performances and how? The number of known auxiliary data points? The training status of targeted model? And the architecture of surrogate model? 
    \item How do the learning regularizations impact the attack performance? Is there any other regularization to boost the performance?
    
    \end{enumerate}

\subsection{Attack Experimental Results}

\subhead{Attack Performance} We first evaluate the attack performance on the three datasets with various sizes and also very small percent known data samples ($<\sim1\%$ out of the whole dataset) correspondingly. \autoref{tab:main_attack} demonstrates the results on the three different datasets (with various sizes and value ranges of labels). From the table, our proposed attack can achieve good performance ($\sim5\%$ or less) while the supervised-learning baseline achieves much worse results in aspects of ALV and AER (the deviation rate of learning labels from the original labels). For instance, even with very few ($1\%$) data samples, our attack achieves a 2.31 ALV/3.47\% AER on the Boston Housing dataset while the baseline obtains 7.54/78\% 
(out of range 5.0 to 50.0). This shows the effectiveness of our attack and the naive semi-supervised learning method could not achieve a better result. This is reasonable since the number of known data samples may not be enough to fine-tune a surrogate model, which cannot be close to the original model. The results on other two datasets also show similar attack results.

\begin{table*}[!h]
\centering
\caption{Comparison of attack performance (ALV and AER) between proposed attack and baseline with different numbers auxiliary known data on Boston Housing price dataset.}
\begin{tabular}{c|cccccc}
\hline
Number & 4 & 6 & 10 & 15 & 20\\\hline
Ours  & 2.31/3.47\% & 1.87/2.95\% &1.43/1.94\% & 1.08/0.73\%  & 0.87/0.65\%\\

Baseline & 7.54/78.31\% & 6.21/62.53\%& 5.12/36.19\%& 4.82/25.42\% & 3.91/15.31\%\\\hline

\hline

\end{tabular}
\label{tab:attack_data}
\end{table*}

\subhead{Impact of Number of Known Labeled Data} Recall that one main regularization for our attack is the semi-supervised learning objective with the auxiliary labeled data samples. We conduct evaluations about the impact of number of known labeled data on the attack performance. For comparison, we also give the results of the baseline method with the same number of known data samples correspondingly. Specifically, we will randomly sample a specific percentage of data samples as the known data from the training set and repeat 5 times. \autoref{tab:attack_data} demonstrates the attack results with different numbers of known labeled data samples. We can observe that the attack performance is getting better as the number of known labeled data samples increases for both methods. However, the semi-supervised learning baseline cannot match with the proposed even with a larger known labeled dataset. This is expected since it is rather difficult for the semi-supervised training to fit all the data with a few data samples. 
\begin{figure}[!h]
	\centering{
		\includegraphics[angle=0, width=0.8\linewidth]{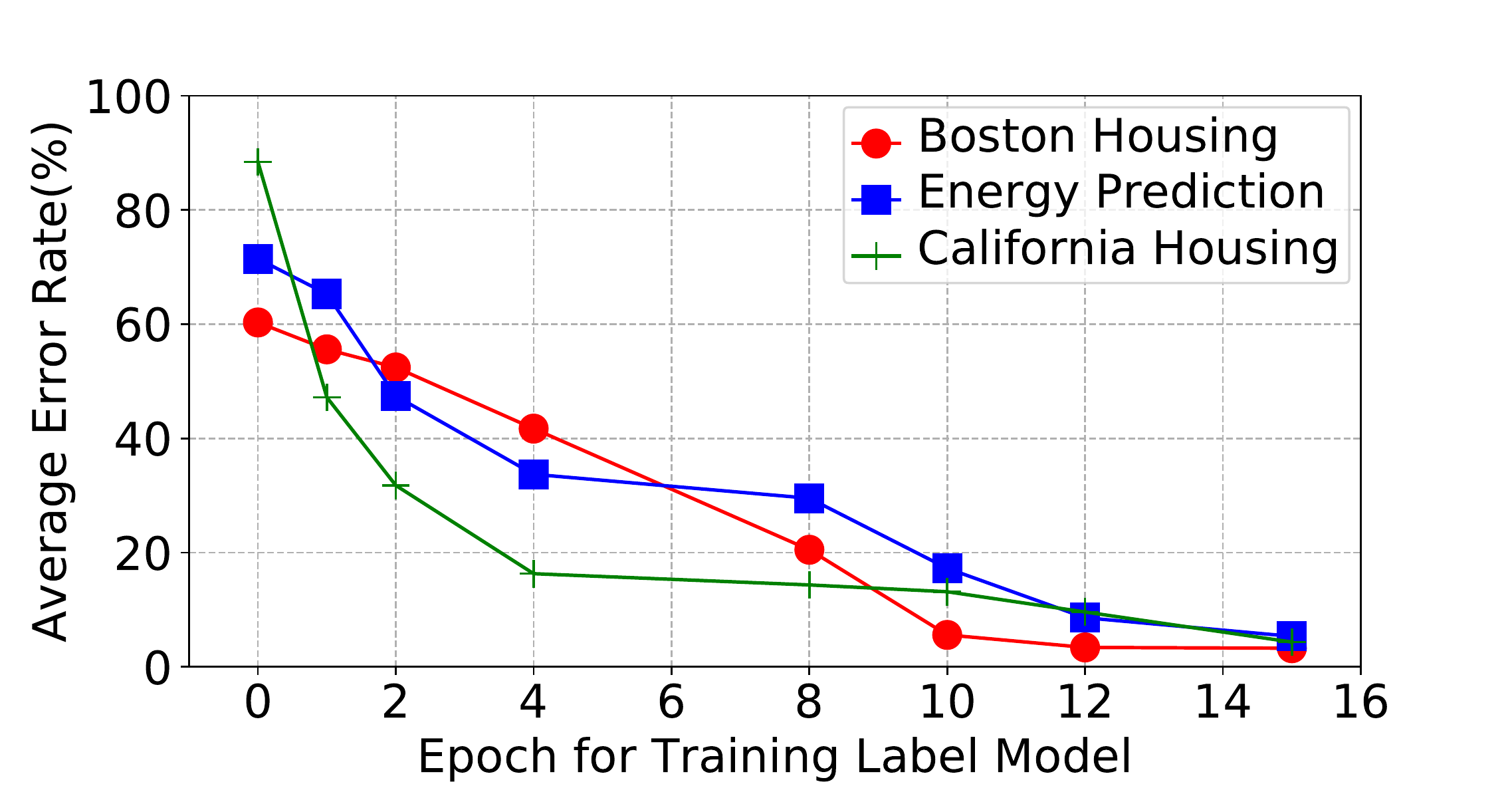}}
	\caption[Optional caption for list of figures]
	{Attack Performance (AER) on Label Model at different training epochs with 3 datasets.}
	\label{fig:att_epoch}
\end{figure}

\subhead{Impact of Training Epoch} We have additionally studied the training status of the target label model. From \autoref{fig:att_epoch}, we can observe the AER of the attack to the untrained model on Boston Housing price dataset decreases as the number of epochs increase, e.g., the starting AER is $60.31\%$ (at training epoch 0), then reduces to $41.72\%$ at (at epoch 4), and to $3.47\%$ at (epoch 15). At the same time, the model performance increases as the model is being to trained to fit the dataset. This is reasonable based on the following two reasons. On the one hand, our attack considers a fixed model at the latter training epochs that the model can fit well with the datasets, i.e., the output prediction score should be close to the ground truth label scores. As the training epochs increases, such distance loss between labels and prediction scores will be closer. On the other hand, it would be easier to train a surrogate model to approximate a over-fitted target model instead of a under-fitted model considering the correct results could be more difficult to learn given the high-dimensional space of neural models could result in more local optimal solutions due to a larger gap.

\subhead{Impact of Surrogate Model} The default setting for the surrogate model in our attack evaluation is the same model of the same architecture as target model but with unknown parameters. For the black-box attack, the attacker may not get the same model architecture as the target model. Thus we evaluate the attack performance with the different surrogate models. For simplicity, we utilize Boston housing price and energy prediction dataset for evaluation. Specifically, we change the architecture of surrogate model from FC3 to FC2 and FC1, and report the attack results in \autoref{tab:att-surr}. From the table, we can observe that the complexity of surrogate model impacts the attack performance. For example, FC-3 surrogate can achieve 3.47\% while FC-1 only achieve 8.12\%. This may due to that the FC-3 has more similarity in model architecture and thus could behave more likely as original label model.

\begin{table}[!ht]
\centering
\caption{Attack performance of proposed attack with different surrogate models.The original label model for both datasets are FC-3.}
\begin{tabular}{c|cccc}
\hline
Surrogate & FC-3 & FC-2 & FC-1 \\\hline
Boston Housing  & 3.47\% & 3.65\% & 8.12\% \\

Energy Prediction & 4.16\% & 4.20\%& 11.27\%\\\hline

\hline

\end{tabular}
\label{tab:att-surr}
\end{table}

\subsection{Ablation Study of Regularization}\label{exp:abl}

We also study the effect of different learning regularizations on the final attack performance. Specifically, we will add/remove the regularizations once a time to check the results and present the corresponding analysis. Table \ref{tab:att-reg} presents the attack results on the Boston Housing price dataset with/without regularizations. From the table, we can observe that both regularization can help to improve attack performance, i.e., AERs/L1 loss getting smaller compared with the baseline (no regularizations). The results are expected. On the one hand, Accuracy loss regularization aims to target the trained model, of which the performance is better (the label and model prediction are smaller). This could greatly help to reduce the search spaces. On the other hand, the core knowledge loss, i.e., semi-supervised learning with auxiliary known datasets can present more precise information for the surrogate model to behave like the target model (considering the target model aims to fit among these data samples). Even with a few number of data samples, such regularization can boost the attack performance. For example, the AER is reduced to 5.16\% from 13.52\% with the knowledge learning regularization. We can observe similar result with the accuracy loss regularization.

\begin{table}[!ht]
\centering
\caption{Attack performance (ALV and AER) with different regularizations on the Boston Housing dataset.}

\begin{tabular}{c|cc}
\hline
Regularizations & No Acc. Loss & Acc. Loss  \\\hline

No Know. Learning  & 3.75/13.52\% & 3.11/8.26\%  \\

Know. Learning& 2.97/5.16\% & 2.31/3.47\%\\\hline

\end{tabular}
\label{tab:att-reg}
\end{table}

\subhead{Triplet Similarity Loss} Besides the two main learning regularizations above utilized in our attack, we have studied another triplet loss regularization \cite{hoffer2015deep,7298682} which may lead to better attack performance. This triplet loss is adopted to control the optimization of labels during the learning. This is based on the observation that the similar cut layer embedding ($\mathcal{E}_i=\mathcal{M}_{user}(x_i)$) can be much likely to output the similar label or close output. 

Specifically, we randomly sample several groups of 3 data samples from one learning attack batch. For instance, a batch of size 5 will get 10 different groups of 3 data samples. For one single group, we denote 3 data samples as $X_s=x_1, x_2, x_3$ with the corresponding cut layer embedding $\mathcal{M}_{user}(x_1), \mathcal{M}_{user}(x_2),\mathcal{M}_{user}(x_3)$. The dummy labels are $y_1^{dummy}, y_2^{dummy}, y_3^{dummy}$. W.l.o.g., we set $x_1$ as the anchor, then the attack triplet loss is defined as below:

\begin{equation}
    L_{triplet}=max(0, \beta+f(X_s))
\end{equation}

where $f(X_s)$ is the distance function among the group of data labels defined as 

\begin{equation}
    \texttt{sign}\{|y_1^{dummy}-y_2^{dummy}|-|y_1^{dummy}-y_3^{dummy}|\}
\end{equation}

The sign $\texttt{sign}$ is $1$ if $Dis(\mathcal{M}_{user}(x_1), \mathcal{M}_{user}(x_2))<Dis(\mathcal{M}_{user}(x_1), \mathcal{M}_{user}(x_3))$ else $-1$. $\beta\geq0$ is the regularization strength parameter. The larger $\beta$ is, the more the triplet loss will be added. We set $\beta=0$ for demonstration. Table \ref{tab:triplet} demonstrates the results with/without triplet regularization on the three datasets respectively. We can observe that the triplet loss regularization helps to improve attack performance very minor. For example, the triplet loss results in 0.05 reduction of L1 loss (0.04\% improvement on AER) on the Boston housing dataset. One possible reason is that the triplet loss could only target a local optimization with a single group of data samples instead of multiple groups among the whole batch of labels, especially the labels with regression are continuous values. It is worth to note that the triplet loss works well with binary classification \cite{7298682}, which is consistent with our results to some extent. We will explore this in the future work, e.g., clustering the cut layer embedding vectors to discrete labels for better triplet loss's optimization.

\begin{table}[!ht]
\centering
\caption{Attack performance (ALV and AER) with/without Triplet similarity loss on the three datasets.}
\begin{tabular}{c|cc}
\hline
Dataset & No Triplet.  & Triplet.  \\\hline

Boston Housing  & 2.31/3.47\% & 2.26/3.23\%  \\
Energy Prediction  & 25.73/5.35\% &25.29/5.24\%  \\
California Housing  & 0.11/4.16\% &0.09/4.09\%  \\
\hline

\end{tabular}
\label{tab:triplet}
\end{table}

%% file: dis.tex
Considering most of attack works are based on the shared gradient information to infer the labels, one popular method is to directly protect the private labels held by the label party. For instance, we can also add noises to the labels to be noisy. Then the attackers can only get the noisy values even they can get the target labels. However, such noise-based protections would jeopardize the training performance to some extent, which is closely related to the amount of added noise. Another alternative method is to obfuscate the gradient information by adding noises \cite{li2022label, zhu2019deep}. That is, before sending the gradients to the user for back-propagation, the label party can process the gradient value with some randomly generated noises under regime of differential privacy \cite{dwork2014algorithmic, dwork2006calibrating,mohammady2020r2dp}, e.g., Laplace or Gaussian noises. We conduct evaluations for both methods correspondingly.

\begin{figure}[!th]
	\centering{
		\includegraphics[angle=0, width=0.8\linewidth]{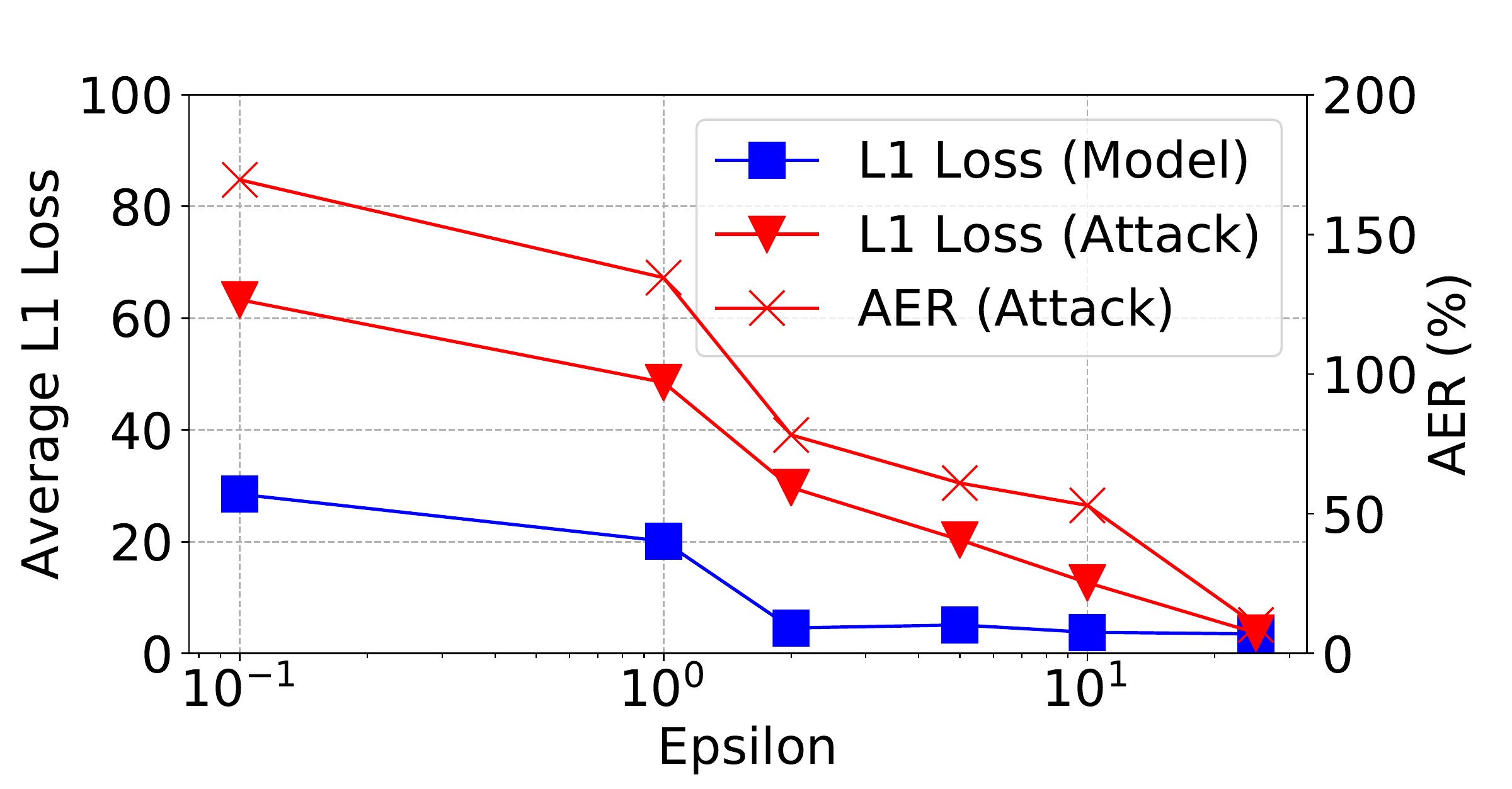}}
	\caption[Optional caption for list of figures]
	{Label defense results on Boston Housing.}
	\label{fig:defense}
\end{figure}

\subhead{Adding Noise to Labels} As discussed above, utilizing noises to obfuscate labels is a commonly used methods. Then we formally define this defense method as following:

\begin{equation}
\widehat{y} \gets y + \xi, 
    ~~~\xi \sim \mathcal{N}
\end{equation}
where the added noises are randomly sampled from some noise distributions $\mathcal{N}$, e.g., Laplace noise.  

\begin{equation}
    \mathcal{N}=Lap(\frac{s}{\epsilon})
\end{equation}
where the sensitivity $s$ is the maximum value of labels. We vary the value of scale $\epsilon=[0.1, 1, 2, 5, 10, 25]$ (the baseline of no protection is $0$). We evaluate the defense scheme with all the dataset and report the attack performance (ALV and AER) and model performance (L1 loss on the test dataset). we repeat every subexperiment 5 times and take the average value. \autoref{fig:defense} report the results. From the figure, we can observe that the attack performance decreases (both AER and L1 loss values increases) while the model performance also decreases accordingly (as the labels have been obfuscated from the original distribution). This demonstrates that the noised label can prevent the leakage. Also this reflects the privacy-utility trade-off. For future work, we can explore more optimized algorithm to add noises to the labels which can preserve both privacy and utility for training.

\subhead{Adding Noise to Gradients} Similar to label obfuscation, we can also add noises to the gradients to protect as following:

\begin{equation}
    \widehat{\nabla_{\mathcal{E}_i}L} \gets \nabla_{\mathcal{E}_i}{L} + \xi, 
    ~~~\xi \sim \mathcal{N}
\end{equation}
where $\nabla_{\mathcal{E}_i}{L}$ is the shared gradient information from label party and $\xi \sim \mathcal{N}$ is the noise also sampled from some distribution. Here we utilize standard Gaussian Distribution \cite{li2022label}, i.e., $\mathcal{N}(0, \sigma)$ and $\sigma=\max(g)/\sqrt{d}$ where $d$ is the dimension of shared gradients. We repeat similar evaluations as label protection and report the results. From \autoref{tab:defend2}, we can observe the similar results to the label defense (lower attack performance and better data protection but lower model training performance). As we add noise to the shared gradients, the attack cannot match a precise gradient as the original one (thus worse attack performance).

\begin{table}[!ht]
\centering
\caption{Gradient Defense results (including both attack and trained model performance) on Boston Housing dataset.}

 \begin{tabular}{c|cc}
\hline
Metric & Attack Performance & Model \\\hline

 No Defense  & 2.31/3.47\% & 2.24  \\

Gradient Defense& 6.72/63.18\% & 6.35\\\hline

\end{tabular}
\label{tab:defend2}
\end{table}

Trade-off between the utility (model performance) and privacy has been always an important topic in the privacy-enhancing technologies. That is, the higher amount of noises could lead to better privacy guarantee but downgrade the model performance. Thus how to wisely add the noise to achieve a better privacy protection without compromising the training performance.